\documentclass[aps,prd,showpacs,twocolumn,notitlepage,nofootinbib,amssymb,amsmath,amsfonts,mathrsfs,
floats,floatfix,superscriptaddress,bibnotes]{revtex4-1}

\usepackage{graphicx}
\usepackage{xcolor}
\usepackage[colorlinks=true]{hyperref}
\hypersetup{colorlinks=true,
            citecolor=blue,
            linkcolor=red,
            urlcolor=red}
\usepackage{multirow,array}
\DeclareMathAlphabet{\pazocal}{OMS}{zplm}{m}{n}
\usepackage{journals}
 
\definecolor{maroon}{cmyk}{0,0.87,0.68,0.32}

\newcommand{\gtwo}{{}_{2}g_{22}}
\newcommand{\gone}{{}_{1}g_{22}}

\begin{document}
\title{Magnetic amplification in premerger neutron stars through resonance-induced magnetorotational instabilities}

\date{\today}

\author{Arthur G. Suvorov}
\email{arthur.suvorov@tat.uni-tuebingen.de}
\affiliation{Theoretical Astrophysics, Eberhard Karls University of T\"ubingen, T\"ubingen 72076, Germany}
\affiliation{Manly Astrophysics, 15/41-42 East Esplanade, Manly, NSW 2095, Australia}

\author{Hao-Jui Kuan}
\affiliation{Max Planck Institute for Gravitational Physics (Albert Einstein Institute), 14476 Potsdam, Germany}

\author{Alexis Reboul-Salze}
\affiliation{Max Planck Institute for Gravitational Physics (Albert Einstein Institute), 14476 Potsdam, Germany}

\author{Kostas D. Kokkotas}
\affiliation{Theoretical Astrophysics, Eberhard Karls University of T\"ubingen, T\"ubingen 72076, Germany}

\begin{abstract}
Tidal resonances in the final seconds of a binary neutron-star inspiral can excite oscillation modes in one or both of the constituents to large amplitudes. Under favorable circumstances, resonant pulsations can overstrain the stellar crust and unleash a torrent of magnetoelastic energy that manifests as a gamma-ray ``precursor flare.'' We show that for realistic, stratified stars rotating with a spin frequency of $\gtrsim30\,$Hz, the fundamental $g$ or its first overtone can also execute a differential rotation in the crust such that a magnetic field of strength $\gtrsim10^{13}\,$G is generated via magnetorotational instabilities. This may help to explain observed precursor rates and their luminosities. Premerger magnetic growth would also provide seed magnetic energy for the postmerger remnant.
\end{abstract}

\maketitle

\section{Introduction}

It is by now beyond doubt that most, if not all, short gamma-ray bursts (SGRBs) are sourced by binary mergers involving at least one neutron star \cite{gold17,marg21}. In rare cases ($\lesssim 5\%$ of events \cite{zhon19}), SGRBs are preceded by ``precursor'' flares up to even $\sim 10\,$s prior to the main event \cite{troj10,zhon19,wang20,copp20}. Given such a delay, these precursors must be somehow launched prior to coalescence.

One model that can successfully explain the observational properties of precursors, such as  their onset times relative to the respective main events, involves the resonant excitation of oscillation modes \cite{tsan12,tsan13,suvo20}. When the orbital frequency inevitably rises to some multiple of the frequency of a natural mode in an inspiraling star, an amount of tidal energy, depending on the ``overlap integral'' \cite{pres77,lai94}, is rapidly siphoned off. For some modes, the resonant amplitude is large enough that the crust becomes overstrained, releasing magnetoelastic energy that fuels a gamma-ray flash. Precursors thus offer a powerful probe of neutron star structure, since the mode spectrum depends on the equation of state (EOS) \cite{kuan22}, magnetic geometry \cite{kuan21b}, rotation rate \cite{kuan23}, nuclear symmetry energy \cite{neil23}, and other microphysical parameters \cite{kuan21,sull23}. 

A potential drawback of the model is that it requires strong magnetic fields, $\boldsymbol{B}$, to explain the observed energies: the maximum magnetically-extractable luminosity from the crust of a star with radius $R$ is given by \cite{tsan12,tsan13}
\begin{equation} \label{eq:maxlum}
L_{\rm prec} \sim 10^{47} \left(\frac{v}{c}\right) \left( \frac {B_{\rm crust}} {10^{13}\, \mbox{G}}\right)^2 \left(\frac{R} {10\, \mbox{km}}\right)^{2} \mbox{~erg~s}^{-1},
\end{equation}
where $v$ is the speed of the mode perturbation.  Reconciling  Eq.~\eqref{eq:maxlum} with observed luminosities (which sometimes reach $\sim 10^{50}\,$erg/s \cite{xiao22,tsan23}) is difficult given that the characteristic inspiral time far exceeds the expected decay timescale(s) in a crust with $B \geq 10^{13}\,$G \cite{gr92,gl21}. If we accept the mode-resonance picture, this poses an astrophysical puzzle: how could $\lesssim5\%$ of GRB-producing mergers contain a star with a magnetar-like field?

Tidal resonances also deposit angular momentum into the stellar interior \cite{lai94}, with the pattern for the resulting angular velocity, $\Omega$, being tied to the mode eigenfunction. As such, one anticipates differentially rotating cavities. In this paper, we show that resonant $g$-modes in realistic, stratified neutron star crusts result in radial angular velocity gradients that are negative, $\partial_{r} \Omega < 0$, over the course of a half mode-period. The crust in one or both stars, some seconds prior to merger when a resonance triggers, thus becomes primed for dynamo activity, most notably the magnetorotational instability (MRI) \cite{ag78,bal91,bal98}.  

The MRI and related instabilities (e.g. Tayler-Spruit dynamo \cite{spruit02,Fuller2019TSdynamo,bar22,barr23}) play an important role in astrophysical systems where differential rotation and `weak' magnetic fields coexist. The most well-studied example is that of accretion disks: the swirling weak fields cause a linear instability that brings about turbulence such that the viscosity can effectively mediate angular momentum exchanges between fluid elements \cite{hgb95,bal98}. The MRI may also be partially responsible for magnetic growth in proto-magnetars \cite{mas07,seigel13,Reboul-Salze2022MRI} and the solar cycles \cite{ag78,kag14}. In the case put forward here for resonance-induced differential rotation, the MRI growth rate is much faster than the mode oscillation period if the star spins at a (pre-resonance) rate of $\nu_{0} \gg 10\,$Hz, and thus the magnetic field can rocket to a saturation value $B_{\rm MRI}$. We estimate $B_{\rm MRI} \gtrsim 3 \times 10^{13} \sqrt{\nu_{0}/30\,\text{Hz}}\,$G for low-order $g$-modes which, in principle, allows Eq.~\eqref{eq:maxlum} to match observed luminosities, alleviating the aforementioned tension between decay and inspiral timescales. 

\section{Resonant $g$-modes}
Microphysical composition or temperature gradients within a fluid allow for buoyancy-restored oscillations ($g$-modes). The characteristic (Brunt-V{\"a}is{\"a}l{\"a}) frequencies of these modes, $N$, depend on the properties of the star through the bulk EOS and the relative `adiabatic' index of perturbative motions. Following Refs.~\cite{kuan21,skk22,kuan23}, we consider the resonant excitation of $g$-modes in the slow reaction limit \cite{pass09} in a stratified star where the spectrum is encoded through a parameter $\delta>0$ (convective stability), defined such that $\Gamma = \gamma(1 + \delta)$ for perturbation, $\Gamma$, and background, $\gamma=(dp/d\epsilon)(\epsilon+p)/\epsilon$, adiabatic exponents. Here $p$ and $\epsilon$ denote the pressure and energy density, respectively, determined via the Tolman-Oppenheimer-Volkoff (TOV) equations over a static background for a specified EOS.

A given $g$-mode, with eigenfunction ${}_{n}\boldsymbol{\xi}_{\ell m}$, is described by three quantum numbers: the overtone number, $n$, and spherical-harmonic indices $\ell$ and $m$. The inertial-frame frequency of such a mode is\footnote{While the tidal potential couples most strongly to $\ell = m =2$ modes in spin-orbit aligned binaries \cite{pres77}, misalignment can lead to sizeable excitations of $m = 1$ modes \cite{kuan23}. We ignore this possibility and concentrate on $\ell = m =2$ here. We also ignore the $\mathcal{O}(\Omega^2)$ centrifugal corrections to the eigenfunction itself.}
\begin{equation} \label{eq:modefreq}
{}_{n}f_{\ell m,i} =  {}_{n}f_{\ell m} - m(1-{}_{n}\mathcal{C}_{\ell}) \nu_{0} - {}_{n}F_{\ell m} ,
\end{equation}
for $\mathcal{C} \approx 0.1$ \cite{krug20,kuan21}, which are EOS-dependent numbers representing leading-order rotational corrections to the static value ${}_{n}f_{\ell m}$. The $F$ terms account for Lorentz and tidal forces, ignored here as they are small unless the binary is very eccentric or $B \gtrsim 10^{15}\,$G \cite{kuan21}. 

\begin{table} 
	\centering
	\caption{Frequencies and maximum amplitudes achieved by tidally-forced $\gone$ ($\gtwo$) modes in equal-mass binaries, with a stratification index of $\delta = 0.005$, for a variety of stellar masses ($M$) and radii ($R$) under the APR4+DH EOS.} \label{tab:modes}
	\begin{tabular}{cccc}
		\hline 
  \hline
		$M$~($M_{\odot}$) & $R$~(km) & Frequency~(Hz) & Max. amp.~$(\times10^{-4})$ \\
		\hline 
		1.20 & 11.34 & 88.46 (59.14) & 2.80 (17.06) \\
    	1.40 & 11.32 & 90.79 (60.61) & 5.69 (12.56) \\
		1.60 & 11.26 & 93.59 (62.32) & 7.81 (10.00) \\
		1.80 & 11.13 & 97.25 (66.48) & 9.13 (7.45) \\
		2.00 & 10.87 & 102.89 (67.84) & 10.39 (5.46) \\
		\hline 
  \hline
	\end{tabular}
\end{table}

We perform a series of computations using a Hamiltonian evolution scheme at 2.5 Post-Newtonian order (i.e. including radiation reaction) to evolve the mode amplitude, $\xi_{\rm amp}(t)$, and binary inspiral simultaneously; see Ref.~\cite{kuan21} for numerical details. While the indices $\gamma$ and $\Gamma$ controlling the $g$-spectra are generally space- and time-dependent, taking a constant approximation for $\delta$ on any given time slice provides accurate spectra to within $\sim5\%$~\cite{kuan22,kuan23}. Tidal heating can raise the temperature-dependent stratification ($\delta \propto T^2$) by a factor $\gtrsim 4$ by $\sim 0.5\,$s prior to merger relative to a `canonical' value of $\delta = 0.005$, anticipated from entropy gradients in a mature star \cite{lai94,xu17}. Note, however, that much larger values have been considered in the literature (e.g. \cite{pass09}). It is useful to remark therefore that the frequency and amplitude scale as ${}_{n}f_{\ell m} \propto \sqrt{\delta}$ and $\xi_{\rm amp, max} \propto \delta^{7/12}$. 

Table~\ref{tab:modes} shows a variety of mode frequencies and resonant amplitudes for the \citet{apr98} (APR4) EOS coupled to a \citet{dh01} (DH) crust, adopted throughout this work as it passes constraints set by GW170817 \cite{abb18} and can support masses consistent with the heaviest (confirmed) neutron star, PSR J0952-0607 ($M_{\rm TOV} \gtrsim 2.1 M_{\odot}$) \cite{rom22}. More precisely, a piecewise-polytropic approximation of the combined EOS is used (see Appendix A of Ref.~\cite{kuan22b}). While this leads to discontinuities in $\gamma(r)$, it greatly simplifies the combined inspiral and mode-excitation evolution and yields quantitatively similar results \cite{kuan21,kuan23,zhao22}.

\subsection{Mode-induced differential rotation} 
\begin{figure}
\includegraphics[width=0.487\textwidth]{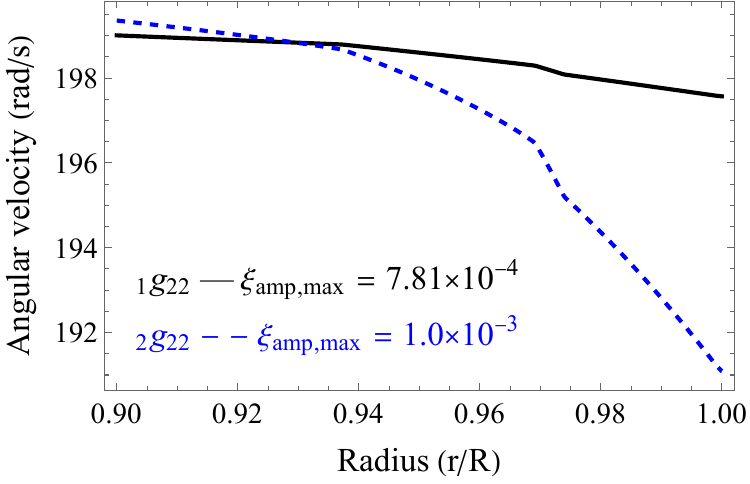}
\caption{Snapshots of the crustal angular velocities, $\Omega(r)$, for $\gone$ (black, solid) and $\gtwo$ (blue, dashed) modes, at the time slice corresponding to the maximum amplitude (Tab.~\ref{tab:modes}), for a star with $\nu_{0} = 100/\pi\,$Hz, $M = 1.6 M_{\odot}$, and $R = 11.26\,$km.}
	\label{fig:omegafig}
\end{figure}

From the evolution scheme detailed above, we can extract the angular velocity imparted to the fluid interior. The $4$-velocity of a perturbation, $\delta u^{\mu}$, associated with a single mode is related to that mode's eigenfunction through 
\begin{equation} \label{eq:difrot}
    \partial_{t} \left({}_{n}\xi_{\ell m}\right)^{\mu} = \delta u^{\mu}/{u^{t}},
\end{equation}
yielding the perturbed angular velocity, viz. $\delta \Omega = \delta u^{\phi}/u^{t}$ assuming that the frame-dragging effect of orbital motion is negligible. Figure~\ref{fig:omegafig} shows the (radial) profile $\Omega = \Omega_{0} + \delta \Omega$ at resonance in the crust for $\gone$ and $\gtwo$ modes, with amplitudes given in Tab.~\ref{tab:modes}, in a $1.6 M_{\odot}$ star rotating with $\Omega_{0} = 200\,$rad/s. Note that, post-resonance, the mode amplitude remains approximately constant over a time-scale of $\lesssim\,$seconds, which should be much longer than that of magnetic amplification (see below). In the case of $\gtwo$ modes, the relative angular velocities between the base and top of the crust (the Rossby number) reaches $\Delta \Omega / \Omega_{0} \sim 0.05$. While $\Delta \Omega / \Omega_{0}$ is lower for $\gone$ modes by a factor $\sim 5$, the shear rate is still large enough that it could instigate amplification, as we now describe.

\section{Magnetorotational instabilities}
Magnetised fluids in a cavity where the angular velocity gradient is sufficiently negative are subject to the MRI, as in the classical example of Keplerian disks \cite{bal91}. The linear growth of the instability, culminating in turbulence, is eventually quenched when either the gradient is erased or magnetic tension can restabilise the system, with the latter implying an episode of rapid magnetic growth. In this study, the cavity in question is a neutron star crust and the source of shear is a tidally-resonant $g$-mode shortly before coalescence (Fig.\,\ref{fig:omegafig}). In the simplest case of ideal magnetohydrodynamics (MHD), MRI activation only requires $\partial_{r} \Omega <0$. In a realistic, stratified medium however, the threshold depends on the microphysical properties of the matter via the Brunt-V{\"a}is{\"a}l{\"a} frequency and the magnetic, $\eta$, and chemical, $\kappa_{\mu}$, diffusivities through $\partial \Omega^2/ \partial \log r<-N^2\eta/\kappa_{\mu}$ \cite{ag78,bal98}. In proto-magnetars with $\gtrsim\,$MeV temperatures or in the radiative interiors of sun-like stars, this criterion is easily satisfied because $\kappa_{\mu}$ exceeds $\eta$ by enormous factors of up to $\sim 10^{14}$ \cite{mir02,mas07}. For a mature neutron star participating in a merger, the situation is less clear and depends on formation history.

Ionic transport simulations can be used to estimate $\kappa_{\mu}$ in a mature crust, treated as a strongly-coupled Coulomb plasma with $\Gamma = Z^2 e^2/ a T \gg 1$ for ion, $Z$, and elementary, $e$, charges, and ion sphere radius $a$. For an accreted crust composed of lighter elements with $\Gamma \gtrsim 200$ (meltdown occurs at $\Gamma \approx 175$ \cite{pan20}), we can infer from Figure 2 in Ref.~\cite{dal12} that $10^{-5} \lesssim \kappa_{\mu}/\omega_{p} a^2 \lesssim 1$ depending on the assumed metallicity, where $\omega_{p}$ is the plasmon frequency. We thus anticipate $10^{-8} \lesssim \kappa_{\mu} \lesssim 10^{-3}$ in CGS units near the base of the crust, increasing weakly as the density drops since $\omega_{p} \propto \rho^{1/2}$ but $a \propto \rho^{-1/3}$ for mass density $\rho$. For a cold, catalyzed crust, diffusion is more easily suppressed and $\kappa_{\mu}$ takes smaller values \cite{hugh11}. 

On the other hand, from Fig.~\ref{fig:omegafig} and standard estimates for the electrical conductivity near $\rho \approx 10^{14}\,\text{g}\,\text{cm}^{-3}$ \cite{cham08},  we have $N^2 \eta/  r| \partial_{r} \Omega^2| \sim 10^{-7} \times (N / 10^{3}~\text{Hz})^{2} (10^{-3} / \xi_{\rm amp})^{2} (T/10^{7}~\text{K})^{\zeta}\,\text{cm}^2/\text{s}$ with $\zeta\approx2$. We therefore expect $\kappa_{\mu} \gg N^2 \eta/  r| \partial_{r} \Omega^2|$ and the MRI to proceed unfettered in a number of astrophysical cases, at least near the crust-core interface where the Brunt-V{\"a}is{\"a}l{\"a} frequency is not too large, especially for recycled pulsars with accreted crusts or at lower temperatures in cases where $\kappa_{\mu}$ falls off slower than $\eta$ (cf. Refs.~\cite{cham08,bez14}). Independently, \citet{fuen23} estimated the Schmidt number -- the ratio of the kinematic viscosity $\nu$ to the diffusivity -- to be of order unity in the crust. This implies a more optimistic scenario for MRI activation, as $\nu \gtrsim 1\,\text{cm}^2\,\text{s}^{-1}$ for $T \lesssim 10^{7}$~K, though again depending on composition. {Such a value for $\nu$ also implies that the dissipative viscous timescale \cite{lai99},
\begin{equation} 
\frac{1}{\tau_{\rm vis}} \sim -\frac{1}{2E} \frac{d E} { d t} \sim \nu \frac{f^2}{v^2}
\end{equation}
for mode energy $E$, is very long compared to the final $\sim$~seconds of inspiral.}

\subsection{Magnetic amplification}
It can be shown that vertical, Alfv{\'e}n-like modes arising from the linearised induction equation with $\delta \boldsymbol{B} \propto e^{i (\boldsymbol{k} \cdot \boldsymbol{x}) - i \omega t}$ in a (electron) fluid where $\kappa_{\mu} \gg N^2 \eta/  r| \partial_{r} \Omega^2|$ have eigenvalues\footnote{Relativistic corrections to this formula have been considered in Refs.~\cite{gam04,yoko05}, though the adjustments are small for low rotation rates unless the star is very compact. We also ignore superfluidity and superconductivity, which could be more important \cite{mend98}.} 
\begin{equation} \label{eq:mri}
\begin{aligned}
0 =&\, \omega^4 - \omega^2 \left[2 (\boldsymbol{k} \cdot \boldsymbol{v}_{\rm A})^2 + \kappa^2 \right]  \\
&+ \left( \boldsymbol{k} \cdot \boldsymbol{v}_{\rm A} \right)^2 \left[(\boldsymbol{k} \cdot \boldsymbol{v}_{\rm A})^2 + \kappa^2 - 4 \Omega^2 \right],
\end{aligned}
\end{equation}
for Alfv{\'e}n velocity $\boldsymbol{v}_{\rm A} = \boldsymbol{B}/\sqrt{4 \pi \rho}$ and epicyclic frequency $\kappa^2 = r^{-3} \partial_{r} (r^4 \Omega^2)$. In this limit, eigenvalues with negative imaginary component exist if $\partial_{r} \Omega < 0$, implying exponentially growing magnetic modes. Eq.~\eqref{eq:mri} shows that the growth rate, $t_{\rm MRI}$, and wavelength, $\lambda_{\rm MRI} = 2\pi/|\boldsymbol{k}|$, of the \emph{fastest growing mode} satisfy 
\begin{equation} \label{eq:tmri}
t_{\rm MRI} \sim \frac{1}{\Omega_{0}}, \qquad \frac{4\pi B^2}{\lambda_{\rm MRI}^2 \rho} = - \left( 1 + \frac{\kappa^2} { 4 \Omega^2}\right) \frac {\partial \Omega^2 } {\partial \log r},
\end{equation}
respectively \cite{bal98,duez06}. Although expressions~\eqref{eq:tmri} are local, it has been argued that the MRI will be globally suppressed in a cavity of extent $\tilde{R}$ if $\lambda_{\text{MRI}} \gtrsim \tilde{R}$ (e.g.~\cite{duez06}), since then the unstable modes will not `fit inside' the cavity. This criterion has been found to match reasonably well to \emph{in situ} MHD simulations where the cavity is some fraction of the equatorial radius of the star \cite{seigel13}. We apply this simple criterion to the crust; full-scale numerical simulations will be conducted in future work. The instability condition $\lambda_{\text{MRI}}(r) \lesssim 0.1R$ translates into \footnote{If the magnetic field strength is larger than the saturation value \eqref{eq:bmri} prior to resonance the inequality is moot.}
\begin{equation} \label{eq:bmri}
B_{\rm MRI} \lesssim \frac{ R \sqrt{\rho} \sqrt{- r \partial_{r} \Omega  \left( 4 \Omega + r \partial_{r} \Omega \right)} }{20 \sqrt{\pi}}.
\end{equation}

Eq.~\eqref{eq:bmri} could change by a factor of a few depending on the exact nature of the MRI, and is strongly sensitive to the eigenfunction through Eq.~\eqref{eq:difrot}. An independent way of estimating the magnetic amplification via the MRI is to assume equipartition between the magnetic, $U_{\rm mag} = B^2/8\pi$, and shear, $U_{\rm shear} = \tfrac{1}{2}(\Delta \Omega/\Omega)^2  (r \Omega)^2 \rho$, energy densities (see, for instance, the discussion in Ref.~\cite{barr23}). 

\begin{figure}
\includegraphics[width=0.487\textwidth]{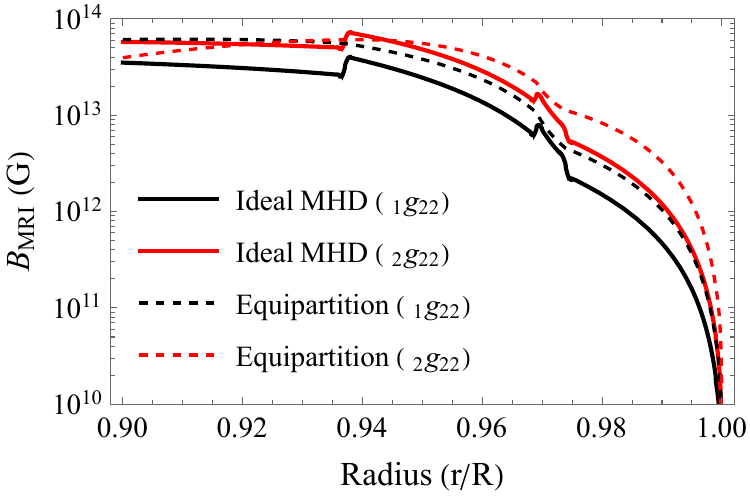}
\caption{Magnetic boosting estimates from the local MHD analysis \eqref{eq:bmri} (solid curves) or from equipartition between the magnetic and shear energies (dashed curves). The same star from Fig.~\ref{fig:omegafig} is used for both $\gone$ (black) and $\gtwo$ (red) modes.}
	\label{fig:bfig}
\end{figure}

Figure \ref{fig:bfig} shows the magnetic fields achieved by $\gone$ and $\gtwo$ modes via either the local MHD analysis \eqref{eq:bmri} or equipartition for the star depicted in Fig.~\ref{fig:omegafig}. The estimates agree within a factor of $\sim 2$, suggesting they are relatively robust. Moreover, the crustal field is large ($\lesssim 10^{14}\,$G) in either case except towards the surface where the density sharply drops to zero. The jagged features visible in the MHD profiles at $r \approx 0.94R$ and $\approx 0.97R$ are artifacts of the jump discontinuities in $\gamma(r)$. For $\gone$ modes, integrating over the radial extent of the crust to estimate the average field from \eqref{eq:bmri}, we find 
\begin{equation} \label{eq:bmri2}
\langle B_{\rm MRI} \rangle \approx 2.1 \times 10^{13}  \left( \frac{\xi_{\rm amp,max}}{10^{-3}} \right)^{1/2} \left( \frac{ \nu_{0}} {30\,\text{Hz}} \right)^{1/2}\,\text{G},
\end{equation}
provided that $\xi_{\rm amp} \ll 1$. The same scaling applies to $\gtwo$ modes but with a slightly larger prefactor, viz. $\approx 3.4 \times 10^{13}\,$G. For moderately fast stars and the resonant amplitudes listed in Tab.~\ref{tab:modes}, amplifications of order \eqref{eq:bmri2} are enough to accommodate the luminosities of precursor flares \eqref{eq:maxlum}, especially if $\nu_{0} \gg 30\,$Hz. For more compact stars, the $\gone$ ($\gtwo$) induced amplifications become larger (smaller), and thus a range of field strengths can be anticipated depending on the binary mass-ratio, spin frequency, stratification, and mode quantum numbers. 

Because MRI activity is limited by the mode oscillation, which reverses the sign of the angular velocity gradient over a half-period, the field may not reach the value \eqref{eq:bmri} if the star is spinning slowly. From Eqs.~\eqref{eq:modefreq} and \eqref{eq:tmri}, the condition $t_{\rm MRI} \lesssim P_{\textrm{mode},i}/2$ reads (for $m=2$)
\begin{equation} \label{eq:minfreq}
\nu_{0} \gtrsim 20.2 \left( \frac {{}_{n}f_{\ell 2}} {100 \text{ Hz}} \right) \text{Hz}.
\end{equation}
In general, we may anticipate a $\gtrsim 2/\pi$ reduction in amplitude applying to \eqref{eq:bmri} if inequality \eqref{eq:minfreq} is only marginally satisfied, with the full amplitude applying in the much-greater-than limit. (Note that $t_{\rm MRI}$ is only a rough estimate and could be somewhat longer in reality \cite{mas16}). 

The above form our main result: tidally-resonant $g$-modes in mature, stratified stars rotating at rates of $\gtrsim 30\,$Hz [Eq.~\eqref{eq:minfreq}] may potentially result in the amplification of the magnetic field to magnetar-like levels [Eq.~\eqref{eq:bmri2}] some $\sim\,$seconds before coalescence. This mechanism has interesting implications for GRB phenomena.

\section{Observational connections} 
During resonance, the magnetic field grows over a timescale of $t_{\rm MRI} \sim 5 \times (30\,\text{Hz}/\nu_{0})\,$ms [Eq.~\eqref{eq:tmri}]. For the overlap integrals pertaining to $g$-modes, this is shorter than the combined time taken for the crust to reach its elastic limit and subsequently emit a flare \cite{tsan12,kuan23}. A $\gtwo$-mode could also incite magnetic growth before a $\gone$-mode actually `breaks' the crust. Therefore, MRI-boosted fields can assist in the launching of a precursor via Alfv{\'e}n waves \cite{tsan13,suvo20}.

The onset time of a given precursor corresponds to an orbital frequency (up to a jet-breakout timescale \cite{suvo20}), allowing one to select from a handful of candidate modes that could become resonant then \cite{tsan12}. The MRI-induced field from a resonant mode, such as those shown in Tab.~\ref{tab:modes}, then leads to a prediction for the precursor luminosity via Eqs.~\eqref{eq:maxlum} and \eqref{eq:bmri2}: the maximum amplitude scales as $\xi_{\rm amp} \propto f_{i}^{-5/6}$ \cite{kuan21}, and thus so does $L_{\rm prec}$. Since stronger stratifications increase $f_{i}$ while spin reduces it, a range of luminosities and onset times can be accommodated by $g$-modes, in principle. For example, the very bright ($\sim 7 \times 10^{49}\,\text{erg s}^{-1}$) precursor to GRB~211211A was observed $\sim1.1\pm0.2\,$s prior to the main event \cite{xiao22}, which matches the resonance-timing criterion for a $\gtwo$ mode in a $M \sim 1.25 M_{\odot}$ star \cite{rast22} provided that $\nu_{0} \approx 69(\sqrt{\delta/0.02} - 1)$Hz (cf. Ref.~\cite{skk22}). If the star is strongly stratified such that $\delta \sim 0.1$, the required spin would be consistent with the observed luminosity (recalling that $\xi_{\rm amp} \propto \delta^{7/12})$, the onset time, and inequality \eqref{eq:minfreq}. 

Magnetic amplification could explain the event rates of precursors. If a relatively large spin frequency \eqref{eq:minfreq} in a stratified star is required to facilitate the MRI (assuming $\kappa_{\mu} \gg \eta$), precursor rarity would be tied to the population of `fast' stars taking part in mergers. Positing some prior distributions for mass, spin, and stratification, we can calculate the probability, $P$, that a binary contains a member for which inequality \eqref{eq:minfreq} is satisfied. Accounting for beaming, we estimate $0.04 \lesssim P \lesssim 0.1$ depending on spin variance, which is in good agreement with the observed precursor rate. We also predict that precursors with $L_{\rm prec} \geq 5 \times 10^{49}\,\text{erg s}^{-1}$ should be rare ($< 1\%$ of events; see Appendix~\ref{sec:appendix} for details).

Many MHD simulations find that stronger seed fields in pre-merging objects result in larger magnetic energies in magnetar remnants \cite{kiuchi14,kiu15,most20}. However, the resolution currently achievable with even state-of-the-art numerical codes may not be enough to globally capture magnetic field amplification from a weak seed field in a proto-magnetar; in reality, the seed energy may have little impact \cite{am22,kiu23}, especially if localised to the crust \cite{rezz23}. A related debate in the literature concerns whether proto-magnetars are capable of collimating a jet that can successfully drill through the polar baryon pollution (see Ref.~\cite{cio20}). In the scenario put forward here, we anticipate that $\gtrsim95\%$ of remnants are formed with weaker seeds (no MRI) and $\lesssim 5\%$ with strong ones (MRI). In light of this, whether the seed field impacts on multi-messenger signals and remnant structure could potentially be tested by observations. For example, if more GRBs with precursors that also go on to display X-ray afterglow plateaus are well modeled by strong field $(>10^{16}$\,G) neutron stars, as for GRBs 080702, 081024 and 100702 \cite{rowl13}, then we could infer that the initial seed affects the merger product in an important way. Possible outliers are GRBs 090510 and 211211A\footnote{The remnant mass for GRB 211211A might have exceeded $\sim 3 M_{\odot}$ \cite{rast22}, favoring instead a black hole.}, though their internal fields could still be ultrastrong \cite{suvk20,suvk21,skk22}. Future observations of gravitational waves will help to clarify the situation \cite{don15}. 

\section{Discussion and caveats}

\begin{figure}
\includegraphics[width=0.487\textwidth]{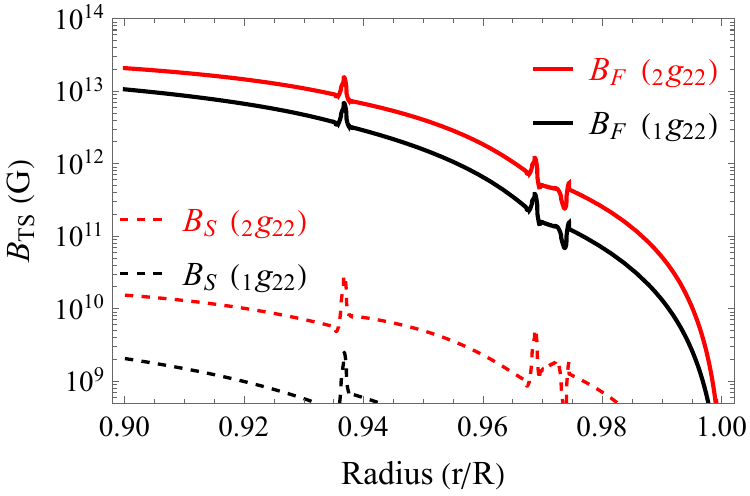}
\caption{Similar to Fig.~\ref{fig:bfig}, though instead for the Tayler-Spruit estimates \eqref{eq:satTS} given in Refs.~\cite{spruit02} (S; dashed) and \protect\cite{Fuller2019TSdynamo} (F; solid).}
	\label{fig:tsfig}
\end{figure}

In this paper, we demonstrate that the resonant excitation of $g$-modes in a binary neutron-star merger induces a negative angular velocity gradient in the crust over the course of a half mode-period, which in turn can excite the MRI if the chemical diffusivity and spin [Eq.~\eqref{eq:minfreq}] are large enough. This instability leads to rapid growth in the crustal field [Eq.~\eqref{eq:bmri}], possibly resolving the mystery of how magnetar-level fields [Eq.~\eqref{eq:maxlum}] seem to be present in mergers releasing precursor flares.

Aside from $g$-modes, other modes have been invoked in the literature to explain precursors. The most notable candidate is the interface mode \citep{tsan12,tsan13}, which exists in stars housing a liquid-to-solid transition at the crust-core (or crust-ocean \cite{sull23}) boundary. Interface modes have a relatively weak azimuthal component however (see, e.g., Figure~4 of Ref.~\cite{sot23}), and thus may not be suitable for introducing large amplifications because the induced shear is small [Eq.~\eqref{eq:difrot}]. Having $B \geq 10^{13}\,$G in a pre-merger flarer may thus require a different explanation -- such as stalled field decay via the Hall attractor \cite{igo18} -- if $i$-modes are the igniters. Other candidates are the intrinsically unstable inertial modes, which also wind up the internal field lines \cite{rezz00,frie16} and may be excited to large amplitudes in rapidly rotating systems \cite{xu17}. The secondary in GW190817 might have been rapidly rotating \cite{bis21}, for example, and so $r$-modes could sometimes play a role.

{An important lesson from the literature is that nonlinear couplings between growing and dormant modes limit saturation amplitudes \cite{arr03,kast10,pnig16}. In the context of large-amplitude core $g$-modes, \citet{wein08} found that three-mode parametric instabilities reduce the peak energy of the dominant parent by factors of $\sim 10^{2}$. Since $E \propto \xi_{\rm amp}^2$ and $\langle B_{\rm MRI} \rangle \propto \xi_{\rm max}^{1/2}$, a leeching of this magnitude would reduce Eq.~\eqref{eq:bmri2} by a factor $\sim 3$. However, the above applies to proto-stars whose $g$-spectra are less dense than that of cold stars, implying a possible underestimate. Nonlinear couplings will be studied in future work; if such a reduction is representative though, the model could not accommodate bright ($L \gtrsim 10^{49} \text{ erg s}^{-1}$) precursors unless $\nu_{0} \gg 20\,$Hz (see Appendix~\ref{sec:appendix}). A related issue is that viscous friction at the crust-core interface could limit crustal amplitudes \cite{lev01}.}

Although we focus on the MRI, it is not the only available option for field amplification in a differentially rotating cavity. Even if strong stratification limits MRI boosting, the Tayler-Spruit dynamo could occur: initially, the toroidal field gets amplified by the winding of the radial field until it becomes unstable to the Tayler instability. When the Tayler-perturbations enter into the non-linear regime a poloidal component is generated, effectively closing the dynamo loop. This mechanism can operate for either positive or negative shear -- unlike the MRI -- and thus could also apply at mode phases where the amplitude is negative. Descriptions for the saturation strength are given by \citet{spruit02} (denoted $B_{\rm S}$) and \citet{Fuller2019TSdynamo} ($B_{\rm F}$), who respectively predict
\begin{equation} \label{eq:satTS}
    B_{\rm S} =  \frac{r\sqrt{4 \pi \rho} \Omega^2 \left(r \partial_{r} \Omega\right)^2}{N^3}, \qquad B_{\rm F} = B_{\rm S} \frac {N^{4/3}} {\left(r \partial_{r} \Omega\right)^{4/3}},  
\end{equation} 
for the radial component. Tayler-Spruit saturation is a complicated process though that depends sensitively on a number of aspects; it is unclear which value applies to a mature crust (see Ref.~\cite{bar22} for a discussion). For completeness, we present both estimates \eqref{eq:satTS} in Figure~\ref{fig:tsfig} for the star from Fig.~\ref{fig:omegafig}. In the optimistic case of the \citet{Fuller2019TSdynamo} estimate, the amplitude is large enough ($\sim 10^{13}\,$G) to be astrophysically relevant since $N \gg r \partial_{r} \Omega$ in the crust for $g$-mode resonances. However, the dynamo may not have sufficient time to amplify the field to $B_{\rm S,F}$ before merger if the initial poloidal field is weak \cite{bar22}. 

Numerical simulations are needed to study such issues more carefully and properly assess whether the MRI or other mechanisms described here operate in Nature. {There are some serious challenges in this direction though in the context of binary merger simulations. (i) Inspiral must begin at a time such that the orbital frequency is less than $\sim 100\,$Hz to capture resonance of low-$n$ $g$-modes. (ii) The crust and thermal gradients, augmenting the $g$-spectrum and determining the activation criterion, must be modelled. (iii) The tidal response for $g$-modes is rather shallow and thus numerical dissipation, which can bias the results, must be treated with special care. It may be more feasible to consider local simulations, along the lines of Ref.~\cite{Reboul-Salze2022MRI}, where some $g$-pattern (e.g., Fig.~\ref{fig:omegafig}) is put in \emph{ad hoc}.} 

\section*{Acknowledgements} AGS acknowledges support from the Alexander von Humboldt Foundation and the European Union's Horizon 2020 Programme under the AHEAD2020 project (Grant No. 871158) provided during the early stages of this work. KDK \& AGS acknowledge support from the HORIZON-MSCA-2022-SE-01 Project: 101131233 — EinsteinWaves.


%


\appendix

\section{Statistical estimations}
\label{sec:appendix}

This material details some statistical calculations regarding MRI activity, including those quoted in the main text. We suppose that the spins of binary neutron-star mergers are distributed according to a Gaussian with mean zero, such that
\begin{equation} \label{eq:gaussian}
f_{\nu}(\chi_{\rm eff},\sigma_{\chi}) = \frac{1}{\sigma_{\chi} \sqrt{2 \pi}} e^{-\frac{\chi_{\rm eff}^2}{2 \sigma_{\chi}^2}},
\end{equation}
where $\chi_{\rm eff}$ is the effective dimensionless spin often used in gravitational-wave studies (e.g.~\cite{zhu18})
\begin{equation}
\chi_{\rm eff} = \frac{ M_{1} \chi_{1} \cos \theta_{1} + M_{2} \chi_{2} \cos \theta_{2}} {M_{1} + M_{2}},
\end{equation}
for component masses $M_{i}$, angles $\theta_{i}$ made between the respective spin and orbital angular momentum vectors, and individual spin magnitudes
\begin{equation}
\chi_{i} = 2 \pi c I_{i} \nu_{i} / G M_{i}^2.
\end{equation}
Here the stellar moments of inertia are represented by $I_{i}$, and $c$ and $G$ denote the speed of light and Newton's constant, respectively. The probability density function defined in Eq.~\eqref{eq:gaussian} matches well to the low-spin prior often used in gravitational-wave data analysis for $\sigma_{\chi} \approx 0.01$, and also to the ``ISO SPIN'' model of \citet{zhu18}, where a uniform distribution for $\cos \theta_{1}$ was considered, if instead $\sigma_{\chi} \approx 0.004$ (see Figure 1 therein). By treating $\sigma_{\chi}$ as a free parameter we can thus study a range of astrophysically-motivated priors.

\subsection{Activation probability}

Since the frequencies and resonant amplitudes of $g$-modes do not scale strongly with the binary mass-ratio $q$ \cite{kuan23}, we assume an equal-mass binary ($M_{1} = M_{2}$) with spin-orbit alignment (expected in a typical astrophysical merger \cite{zhu18}) for simplicity. The observed Galactic binary neutron-star mass distribution, $f_{M}$, is also well-modeled by a Gaussian of mean $1.33 M_{\odot}$ and standard deviation $\sigma_{M} \approx 0.11 M_{\odot}$ \cite{kiz13}, which we adopt here. We further assume a distribution of stratifications, $f_{\delta}$, that runs uniformly from $\delta_{\rm min} = 0$ to $\delta_{\rm max} = 0.2$, the latter value corresponding to the largest values considered by \citet{pass09}. Each star is assumed to have the same stratification (i.e. EOS and temperature). With these simplifying but reasonable assumptions, we can calculate the probability that there will be a member of the binary such that inequality \eqref{eq:bmri2} is satisfied, viz.
\begin{equation} \label{eq:prob}
\begin{aligned}
P& \propto  \varepsilon \int d \nu_{1} d \nu_{2} d M d \delta f_{\nu}(\chi_{\rm eff}, \sigma_{\chi}) f_{\delta}(\delta) f_{M}(M) \\
&\times \mathbf{1} \left[ \nu_{i} > 20.2 \left(\frac {{{}_{1}f_{2 2}}} {100 \text{Hz}} \right) || \nu_{i} > 20.2 \left(\frac {{{}_{2}f_{2 2}}} {100 \text{Hz}} \right)  \right],
\end{aligned}
\end{equation}
normalized by the total value without the conditionals, where the indicator function $\mathbf{1}$ restricts the integral to the appropriate domain. In the above, the factor $\varepsilon \leq 1$ accounts for beaming. While precursor flares are expected to be quasi-isotropic \cite{tsan12}, we posit a factor $\lesssim 2$ reduction due to beaming based on the $\ell = 2$ mode pattern. By building a high-resolution table of $g$-mode frequencies and amplitudes as a function of $M$ and $\delta$ (essentially extending Tab.~\ref{tab:modes}), we evaluate expression \eqref{eq:prob} using a Monte-Carlo scheme (with sampling such that the expected error is $\lesssim 2\%$). We obtain an empirical fit of the form
\begin{equation} \label{eq:probfit}
P \approx 0.04 \times  \left(\frac{ \varepsilon} {0.5} \right) \left( \frac{\sigma_{\chi}} {0.005} \right)^{1.9},
\end{equation}
which matches well with the observed rate of SGRB precursors. However, varying ratios of $\kappa_{\mu}/\eta$ could also contribute to the rarity of precursor observations: the MRI may not activate if $\delta$ (i.e. $N^2$) is too large or diffusion outpaces mode growth. For better or worse, the magnetic diffusivity $\eta$ scales strongly with temperature, density, and the crustal impurity concentration, thereby varying by many orders of magnitude from star to star depending on the cooling and accretion history (see Section 9.3 in Ref.~\cite{cham08}). It is therefore difficult to formally estimate the probability that $\kappa_{\mu} \gg N^2 \eta/  r| \partial_{r} \Omega^2|$ is satisfied, but equation \eqref{eq:probfit} could easily be reduced by a further factor $\gtrsim 2$ depending on the binary formation channel, as discussed in the main text. The final result may also change by a factor of order unity had we used a different EOS. It is thus not difficult to match the $\lesssim 5\%$ observation rate if $\sigma_{\chi} \gtrsim 0.01$.

\subsection{Magnetic growth}

Using equation \eqref{eq:mri}, we can go a bit further and also calculate, assuming the MRI activates, cumulative distribution functions (CDFs) for the magnetic field strengths. Owing to the complexity of the resulting integrals, we simplify the calculation by assuming that one component of the binary is static. More precisely, we compute
\begin{equation} \label{eq:fz}
\begin{aligned}
F_{n}(z) \propto& \int d \nu d M d \delta f_{\nu}(\chi, \sigma_{\chi}) f_{\delta}(\delta) f_{M}(M) \\
&\times \textbf{1}(z > B_{\rm MRI}) \textbf{1} \left[ \nu > 20.2 \left(\frac {{}_{n}f_{2 2}} {100 \text{Hz}} \right) \right].
\end{aligned}
\end{equation} 
The above integral \eqref{eq:fz} defines the proportion of MRI-activating binaries that generate magnetic field strengths $z > B$ for either $n=1$ or $n=2$ modes.

\begin{figure}[ht]
\begin{center}
\includegraphics[width=0.497\textwidth]{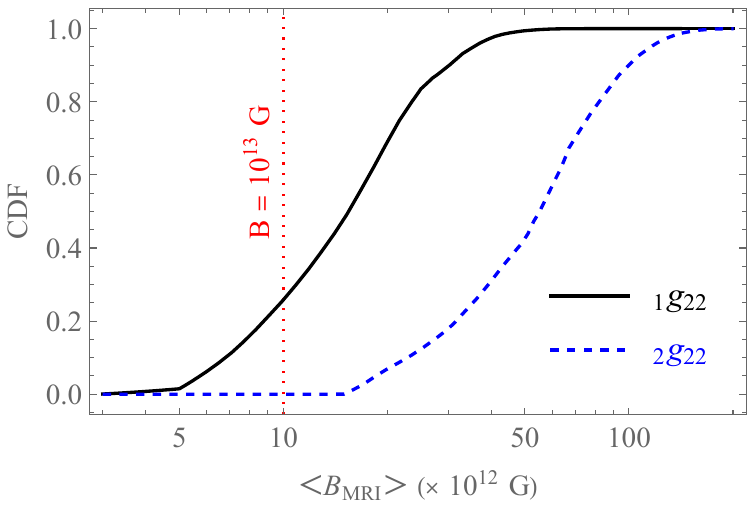}
\end{center}
\caption{CDFs for crustal magnetic field strengths achieved by the MRI (assuming it activates), for either $\gone$ (black, solid) or $\gtwo$ (blue, dashed) modes and $\sigma_{\chi} = 0.01$.}
\label{fig:mag_cdfs}
\end{figure}

The resulting CDFs are shown in Figure \ref{fig:mag_cdfs} assuming $\sigma_{\chi} = 0.01$, as appropriate for the low-spin prior. From these curves we anticipate that most events induce a saturation magnetic field strength that exceeds $10^{13}$~G. For $\gtwo$-modes, the resonant amplitude exceeds $10^{-3}$ in $> 99\%$ of cases (Tab.~\ref{tab:modes}) for the mass prior we have adopted, and thus a negligible portion of activations result in fields below $10^{13}$~G (dotted line). The tendancy for $\gtwo$ modes to induce larger magnetic growths (with some $\sim 10\%$ exceeding $10^{14}$~G) is furthered by the fact that the mode frequency is generally lower, meaning that greater spins are implied if the MRI is to activate (which is rarer than for $\gone$ modes), and hence larger $\langle B_{\rm MRI} \rangle$ values are achieved overall. Had we used a more (less) optimistic spin prior, the graphs would skew further to the right (left). In a similar vein, if we capped the stratification index to a smaller value (or assumed it clustered around a smaller value) and/or postulated reductions due to nonlinear couplings, the distribution would skew towards the left. A more thorough statistical investigation considering these aspects will be carried out elsewhere.

\subsection{Luminosities}

\begin{figure}[ht]
\begin{center}
\includegraphics[width=0.497\textwidth]{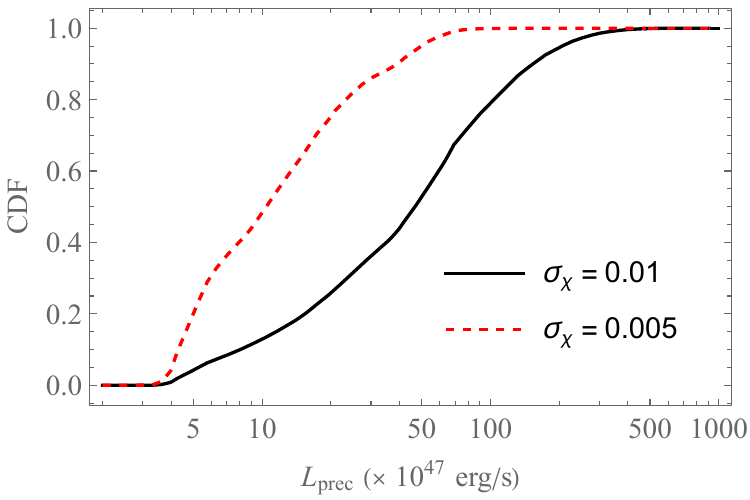}
\end{center}
\caption{Precursor luminosity CDFs for $\sigma_{\chi} = 0.01$ (black, solid) and $\sigma_{\chi} = 0.005$ (red, dashed). Magnetic amplification from both $\gone$ and $\gtwo$ modes are considered.}
\label{fig:lum_cdfs}
\end{figure}

To compare with precursor observations more directly we can also compute CDFs for the maximum magnetically-extractable luminosity \eqref{eq:maxlum}. The relevant integral is similar to that of \eqref{eq:fz}, with the argument of the indicator function replaced by the luminosity in the obvious way. The results are shown in Figure \ref{fig:lum_cdfs}. Even in the more optimistic case of $\sigma_{\chi} = 0.01$, we anticipate that very bright ($L_{\rm prec} > 5 \times 10^{49} \text{ erg s}^{-1}$) precursors should be rare, constituting $\sim 1$ in $500$ events. More typical luminosities should be those that cluster around the expected value, $L_{\rm avg} \approx 6 \times 10^{48} \text{ erg s}^{-1}$, which is in broad agreement with precursor observations. If, however, the precursor-inducing population had a lower spin variance or Eq.~\eqref{eq:bmri2} reduces significantly in the face of nonlinearity, it would be difficult for MRI boosting to explain events such as GRB 211211A, whose peak luminosity reached $\approx 7 \times 10^{49} \text{ erg s}^{-1}$ \cite{xiao22}. For $\sigma_{\chi} = 0.005$ we find in fact that only one in $\sim 10^{10}$ events could reach $L_{\rm prec} > 5 \times 10^{49} \text{ erg s}^{-1}$ (i.e., another mechanism would be necessary to explain at least this particular precursor).

\end{document}